# Thermal circuits assembling and state-space extraction for modelling heat transfer in buildings

Christian Ghiaus, Naveed Ahmad

Univ Lyon, CNRS, INSA-Lyon, Université Claude Bernard Lyon 1, CETHIL, UMR5008, F-69621, Villeurbanne, France

## Abstract

State-space representation is essential in the theory of dynamic systems. This paper introduces a methodology for obtaining state-space representation from the thermal models of elementary components of a building by the conjunction of two methods: 1) assembling of thermal circuits and 2) state-space extraction from thermal circuit. These methods are fully illustrated on a very simple model and tested on a real house of about 100 m$^2$ on which detailed measurements were achieved for 40 days at a time step of 10 min. The errors obtained between the measurements and the simulation results are in the order of $\pm 1$ °C for a single zone and $\pm 2$ °C for seven thermal zones. Besides simulation, parameter identification and control, the methods for assembling thermal circuits and extraction of state-space representation may be useful in Building Information Modelling (BIM).

## 1  Introduction

Thermal networks are graphic representations of systems of differential algebraic equations (DAE) which model heat transfer by conduction (described by the weak form of the heat equation [1]), convection (described by Newton law) and long wave radiative exchange (described by using radiosity [2]). The resistor-capacitor (RC) models have physical meaning that allows the evaluation of the modelling hypothesis considered for buildings and their urban environment [3].

Thermal networks are used for defining the models of elementary components (e.g. walls, floors, doors, windows, etc.) in energy balance method [4, 5]. Energy balance method is the recommended method of ASHRAE [6] and the basis of the CEN standard for calculation of the design heat load [7] as well as of other CEN standards related to thermal performance of buildings [8, 9]. Other procedures are seen as variants or simplifications of the heat balance method. In this method, the set of equations is integrated numerically, generally by using existing solvers. It is the case of many commercial simulation software [10] such as TRNSYS [11], EnergyPlus [12], IDA ICE [13], of research oriented tools, like ESP-r [14], CODYRUN [15] or of tools developed "in house" [16, 17, 18, 19, 20]. Another approach is to use equation-based modeling. In this case, the computational causality (i.e. the input – output relation) is defined after the model was constructed and can be changed. Then a simulation engine performs the calculations [21, 22, 19]. This has the advantage of using the same model for different sets of inputs and outputs but can generate ill-posed problems [4].

State-space is the most used input-output representation of linear models, as shown by recent reviews on modelling of building energy systems [23, 14] and on strategies for building energy management [24, 25]. State-space representation is widely used for model identification and calibration. In model identification, the structure of the model is proposed and then the parameters are identified by minimizing the error between the output of the model and the measured data. An essential issue is the structure of the model on which the experimental data is projected. This is done mostly empirically by using models with a variable number of states [26, 27, 28, 29, 30]. The procedure of model calibration is very similar to parameter identification: use optimisation techniques to fit the model to data by changing the values of the parameters. The main difference is that the parameters obtained by calibration of physical models have physical significance [31, 32, 33, 34].

State-space representation is widely used for model order reduction, which can be done numerically, when the state-space model is known, as is the case for walls [35] or by projecting the results obtained by simulation on a given structure [36]. A key point in model order reduction is the model order selection [37, 38].

State-space model is the most used representation in modern control theory. One approach to obtain the state-space representation is to use a thermal network for the model of the building [39] and to identify the parameters of the state-space representation from input-output data [40, 41, 42, 43]. Usually, the model used for controller synthesis has one state variable: the indoor temperature [44, 45, 46, 47, 48, 49, 50], although state-space models were obtained from the thermal network of a room for 4 states [51], 6 states [52] or for 17 states [53].

Since thermal networks are widely used for modelling heat transfer and state-space is the most used representation in control theory, state-space extraction from thermal networks is of the highest interest. A solution to obtain systematically the state-space representation is by using nodal/mesh analysis to reduce the number of undesired variables [54]. Another state-space extraction method uses the concepts of tree (a sub-graph of the original graph containing no loops) and co-tree (a sub-graph of the original graph containing the edges removed to form the tree) but it requires symbolic manipulation [55]. Commercial implementations of the state-space extraction are not documented [56].

This paper introduces a method for obtaining state-space models for complex heat transfer in buildings by extracting them from thermal circuits. The thermal circuits are obtained by assembling elementary models which represent building elements (e.g. walls, windows, ventilation, etc.).

## 2 Obtaining the system of differential-algebraic of equations (DAE) from the thermal networks

A thermal circuit is a weighted directed graph in which the branches represent heat flow rates crossing thermal resistances and thermal sources while the nodes represent temperatures to which thermal capacities and heat flow rate sources are connected [57, 58].

### 2.1 Algebraic description of the thermal circuits
A thermal circuit is described by three matrices:

1) oriented incidence matrix **A** with the number of lines equal to the number of branches and the number of columns equal to the number of nodes:

$$a_{ij} = \begin{cases} 0 & \text{if branch } i \text{ is not connected to the node } j \\ -1 & \text{if branch } i \text{ leaves the node } j \\ 1 & \text{if branch } i \text{ enters the node } j \end{cases} \quad (1)$$

2) diagonal matrix **G** of conductances of dimension equal to the number of rows of **A**:

$$g_{ij} = \begin{cases} R_i^{-1} & \text{for } i = j \\ 0 & \text{for } i \neq j \end{cases} \quad (2)$$

3) diagonal matrix **C** of capacitances of dimension equal to the number of columns of **A**:

$$c_{ij} = \begin{cases} C_i & \text{for } i = j \\ 0 & \text{for } i \neq j \end{cases} \quad (3)$$

and three vectors with elements 0 and 1:
1) vector **b** of temperature sources of dimension equal to the number of rows of **A**, with the elements

$$b_i = \begin{cases} 1 & \text{for temperature source on branch } i \\ 0 & \text{otherwise} \end{cases} \quad (4)$$

2) vector **f** of flow sources of dimension equal to the number of columns of **A**, with the elements

$$f_i = \begin{cases} 1 & \text{for flow source in node } i \\ 0 & \text{otherwise} \end{cases} \quad (5)$$

3) vector **y** of temperature outputs of dimension equal to the number of columns of **A**, pointing to the temperature nodes that are used as outputs:

$$y_i = \begin{cases} 1 \text{ for temperature as output variable} \\ 0 \text{ otherwise} \end{cases} \quad (6)$$

Then, each thermal circuit (TC) is described by the set of arrays:

$$TC = \{\mathbf{A}, \mathbf{G}, \mathbf{b}, \mathbf{C}, \mathbf{f}, \mathbf{y}\}. \quad (7)$$

## 2.2 System of differential-algebraic equations (DAE)

By writing the differences of temperature,

$$\mathbf{e} = -\mathbf{A}\,\boldsymbol{\theta} + \mathbf{b} \quad (8)$$

the balance of heat rates in nodes,

$$\mathbf{C}\,\dot{\boldsymbol{\theta}} = \mathbf{A}^T\,\mathbf{q} + \mathbf{f} \quad (9)$$

and the constitutive laws for heat transfer,

$$\mathbf{q} = \mathbf{G}\,\mathbf{e} \quad (10)$$

where $\mathbf{q}$ is the vector of heat flow rates, and by substituting $\mathbf{e}$ from equation (8) into equation (10), we obtain Karush-Kuhn-Tucker (KKT) equations [59]:

$$\begin{cases} \mathbf{G}^{-1}\mathbf{q} + \mathbf{A}\boldsymbol{\theta} = \mathbf{b} \\ -\mathbf{A}^T\mathbf{q} + \mathbf{C}\dot{\boldsymbol{\theta}} = \mathbf{f} \end{cases} \quad (11)$$

By eliminating $\mathbf{q}$ from (11), we obtain the set of differential-algebraic algebraic equations describing the thermal circuit:

$$\mathbf{C}\,\dot{\boldsymbol{\theta}} = -\mathbf{A}^T\,\mathbf{G}\,\mathbf{A}\,\boldsymbol{\theta} + \mathbf{A}^T\,\mathbf{G}\,\mathbf{b} + \mathbf{f} \quad (12)$$

# 3 Assembling the thermal circuits

It is easy and convenient to obtain thermal circuits for different elements of the building (walls, floors, windows, doors, etc.). Then, the model of a whole building may be obtained by assembling the elements. Assembling is different from coupling. In coupling, the models of the elements form a set of equations which is solved numerically; in assembling, the model of the whole building is obtained first and then the system of equations is solved. The advantage of assembling is that the model can be analysed: the eigenvalues and the eigenvectors of the whole system can be obtained.

The problem of circuit assembling is to obtain the thermal circuit $TC$ by knowing that some nodes of the elementary circuits $TC_1, TC_2, \ldots, TC_n$ are common to several circuits. Since a thermal circuit is described by the set of arrays, $TC = \{\mathbf{A}, \mathbf{G}, \mathbf{b}, \mathbf{C}, \mathbf{f}, \mathbf{y}\}$, the aim of assembling is to form the global KKT system:

$$\begin{bmatrix} \mathbf{G}^{-1} & \mathbf{A} \\ -\mathbf{A}^T & \mathbf{C}s \end{bmatrix} \begin{bmatrix} \mathbf{q} \\ \mathbf{\theta} \end{bmatrix} = \begin{bmatrix} \mathbf{b} \\ \mathbf{f} \end{bmatrix} \tag{13}$$

or, by using the notations:

$$\mathbf{K} \equiv \begin{bmatrix} \mathbf{G}^{-1} & \mathbf{A} \\ -\mathbf{A}^T & \mathbf{C}s \end{bmatrix} ; \mathbf{u} \equiv \begin{bmatrix} \mathbf{q} \\ \mathbf{\theta} \end{bmatrix} ; \mathbf{a} \equiv \begin{bmatrix} \mathbf{b} \\ \mathbf{f} \end{bmatrix} \tag{14}$$

to form the equation:

$$\mathbf{K}\mathbf{u} = \mathbf{a} \tag{15}$$

from the KKT models of the elementary systems (walls, floors, doors, windows, etc.):

$$\begin{bmatrix} \mathbf{G}_i^{-1} & \mathbf{A}_i \\ -\mathbf{A}_i^T & \mathbf{C}_i s \end{bmatrix} \begin{bmatrix} \mathbf{q}_i \\ \mathbf{\theta}_i \end{bmatrix} = \begin{bmatrix} \mathbf{b}_i \\ \mathbf{f}_i \end{bmatrix} \tag{16}$$

We can write equation (16) as

$$\mathbf{K}_i \mathbf{u}_i = \mathbf{a}_i \tag{17}$$

where:

$$\mathbf{K}_i \equiv \begin{bmatrix} \mathbf{G}_i^{-1} & \mathbf{A}_i \\ -\mathbf{A}_i^T & \mathbf{C}_i s \end{bmatrix} ; \mathbf{u}_i \equiv \begin{bmatrix} \mathbf{q}_i \\ \mathbf{\theta}_i \end{bmatrix} ; \mathbf{a}_i \equiv \begin{bmatrix} \mathbf{b}_i \\ \mathbf{f}_i \end{bmatrix} \tag{18}$$

Let's note the dissembled block matrix $\mathbf{K}_d$ and the disassembled block vectors $\mathbf{u}_d$, $\mathbf{a}_d$, the matrix and the vectors obtained by simply placing in order the matrices and the vectors of the elementary models described by equation (17):

$$\mathbf{K}_d \equiv \begin{bmatrix} \mathbf{K}_1 & \cdots & \mathbf{0} \\ \vdots & \ddots & \vdots \\ \mathbf{0} & \cdots & \mathbf{K}_n \end{bmatrix} \mathbf{u}_d \equiv \begin{bmatrix} \mathbf{u}_1 \\ \vdots \\ \mathbf{u}_n \end{bmatrix} ; \mathbf{a}_d \equiv \begin{bmatrix} \mathbf{a}_1 \\ \vdots \\ \mathbf{a}_n \end{bmatrix} \tag{19}$$

There is a disassembling matrix $\mathbf{A}_d$ which transforms the assembled vectors $\mathbf{a}$ and $\mathbf{u}$ into the dissembled vectors $\mathbf{a}_d$ and $\mathbf{u}_d$:

$$\mathbf{a}_d = \mathbf{A}_d \mathbf{a}; \quad \mathbf{u}_d = \mathbf{A}_d \mathbf{u}; \tag{20}$$

The relations between the global and the elementary matrices and vectors are:

$$\mathbf{K} = \mathbf{A}_d^T \mathbf{K}_d \mathbf{A}_d \tag{21}$$

$$\mathbf{u} = \mathbf{A}_d^T \mathbf{u}_d \tag{22}$$

$$\mathbf{a} = \mathbf{A}_d^T \mathbf{a}_d \tag{23}$$

The elements of the assembled circuit, $TC = \{\{\mathbf{A}, \mathbf{G}, \mathbf{b}, \mathbf{C}, \mathbf{f}, \mathbf{y}\}$, are then obtained from the partition of the arrays:

$$\mathbf{K} = \begin{bmatrix} \mathbf{G}^{-1} & \mathbf{A} \\ -\mathbf{A}^T & \mathbf{C}s \end{bmatrix}; \quad \mathbf{u} = \begin{bmatrix} \mathbf{q} \\ \boldsymbol{\theta} \end{bmatrix}; \quad \mathbf{a} = \begin{bmatrix} \mathbf{b} \\ \mathbf{f} \end{bmatrix} \tag{24}$$

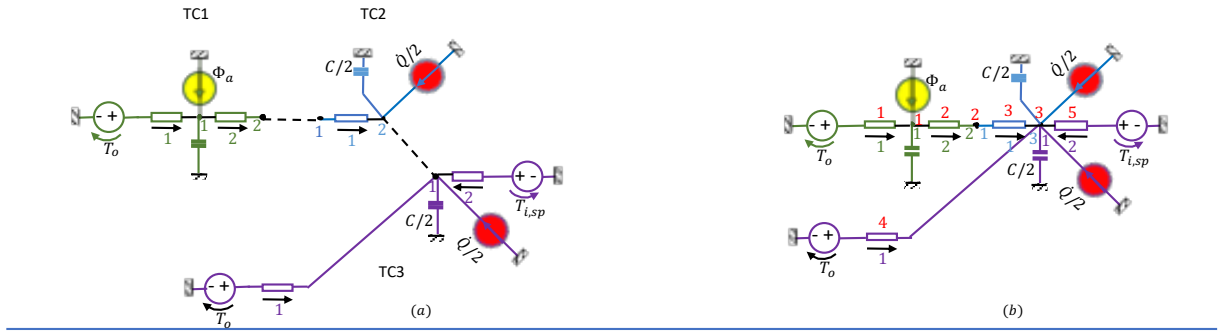

Figure 1 Simple example for the assembling of thermal circuits: a) three disassembled circuits with local indexing of nodes and branches (the dashed lines show the nodes which are in common); b) the assembled circuit with local and global indexing of nodes and branches; c) the matrices and the vectors characterising the disassembled thermal circuits; d) the disassembling matrix and the transformation of the assembled vector into disassembled vectors; e) the matrices and the vectors characterising the assembled thermal circuit.

An illustrative example is given in Figure 1. There are three circuits:
- TC1 with two branches and two nodes,
- TC2 with one branch and two nodes, and
- TC3 with two branches and one node.

The second node of TC1 is put in common with the first node of TC2 and the second node of TC2 is put in common with the first node of TC3 (Figure 1a). The local indexes of the dissembled circuits correspond to global indexes in the assembled circuit (Figure 1b). Each disassembled circuit is characterised by the set of arrays $TC_i = \{\mathbf{A}_i, \mathbf{G}_i, \mathbf{b}_i, \mathbf{C}_i, \mathbf{f}_i, \mathbf{y}_i\}$, where $i = 1, 2, 3$, with values given by equations (1)…(5) (Figure 1c). The disassembling matrix, $\mathbf{A}_d$, transforms the assembled variables, $\mathbf{u}$, into disassembled variables, $\mathbf{u}_d$ (Figure 1d). Finally, the assembled circuit is characterised by a set of arrays, $TC = \{\mathbf{A}, \mathbf{G}, \mathbf{b}, \mathbf{C}, \mathbf{f}, \mathbf{y}\}$ (Figure 1e).

## 4 Extract state-space model from thermal circuits

If the thermal circuit contains nodes without heat capacity, the matrix $\mathbf{C}$ is singular. In order to obtain the state-space model, the equations corresponding to the nodes without heat capacity need to be eliminated from the system of equations (12) [4]. By partitioning the matrix $\mathbf{C}$,

$$\mathbf{C} = \begin{bmatrix} \mathbf{0} & \mathbf{0} \\ \mathbf{0} & \mathbf{C}_C \end{bmatrix} \tag{25}$$

where $\mathbf{C}_C$ corresponds to the nodes having capacities, the set of equations (12) may be written as:

$$\begin{bmatrix} \mathbf{0} & \mathbf{0} \\ \mathbf{0} & \mathbf{C}_C \end{bmatrix} \begin{bmatrix} \dot{\boldsymbol{\theta}}_0 \\ \dot{\boldsymbol{\theta}}_C \end{bmatrix} = \begin{bmatrix} \mathbf{K}_{11} & \mathbf{K}_{12} \\ \mathbf{K}_{21} & \mathbf{K}_{22} \end{bmatrix} \begin{bmatrix} \boldsymbol{\theta}_0 \\ \boldsymbol{\theta}_C \end{bmatrix} + \begin{bmatrix} \mathbf{K}_{b1} \\ \mathbf{K}_{b1} \end{bmatrix} \mathbf{b} + \begin{bmatrix} \mathbf{I}_{11} & \mathbf{0} \\ \mathbf{0} & \mathbf{I}_{22} \end{bmatrix} \begin{bmatrix} \mathbf{f}_0 \\ \mathbf{f}_C \end{bmatrix} \tag{26}$$

where:
- $\boldsymbol{\theta}_0$ and $\mathbf{f}_0$ correspond to the nodes without thermal capacity;
- $\boldsymbol{\theta}_C$ and $\mathbf{f}_C$ correspond to the nodes with thermal capacity;
- $\mathbf{C}_C$ is the bloc of the partitioned matrix $\mathbf{C}$ for which the elements on the diagonal are non-zero;
- $\mathbf{K}_{11}, \mathbf{K}_{12}, \mathbf{K}_{21}$, and $\mathbf{K}_{22}$ are blocs of the partitioned matrix $\mathbf{K}$ obtained according to the partioning of the matrix $\mathbf{C}$;
- $\mathbf{K}_{b1}$ and $\mathbf{K}_{b2}$ are blocs of the partitioned matrix $\mathbf{K}_b$ obtained according to the partitioning of the matrix $\mathbf{C}$;
- $\mathbf{I}_{11}$ and $\mathbf{I}_{22}$ are identity matrices.

The state equation of the state-space model is:

$$\dot{\boldsymbol{\theta}}_C = \mathbf{A}_S \boldsymbol{\theta}_C + \mathbf{B}_S \mathbf{u} \tag{27}$$

where the state matrix is

$$\mathbf{A}_S = \mathbf{C}_C^{-1}(-\mathbf{K}_{21}\mathbf{K}_{11}^{-1}\mathbf{K}_{12} + \mathbf{K}_{22}) \qquad (28)$$

and the input matrix is

$$\mathbf{B}_S = \mathbf{C}_C^{-1}[-\mathbf{K}_{21}\mathbf{K}_{11}^{-1}\mathbf{K}_{b1} + \mathbf{K}_{b2} \quad -\mathbf{K}_{21}\mathbf{K}_{11}^{-1} \quad \mathbf{I}] \qquad (29)$$

If the outputs are temperatures of nodes with capacities, the observation matrix $\mathbf{C}_S$ extracts their values from the state vector and the feed-through matrix, $\mathbf{D}_S = \mathbf{0}$. If the outputs are temperatures from nodes without capacities, the observation equation can be obtained from the first row of equation (26) [4]:

$$\begin{aligned}\boldsymbol{\theta}_0 &= -\mathbf{K}_{11}^{-1}(\mathbf{K}_{12}\boldsymbol{\theta}_C + \mathbf{K}_{b1}\mathbf{b} + \mathbf{I}_{11}\mathbf{f}_0) \\ &= -\mathbf{K}_{11}^{-1}\left(\mathbf{K}_{12}\boldsymbol{\theta}_C + [\mathbf{K}_{b1} \quad \mathbf{I}_{11} \quad \mathbf{0}]\begin{bmatrix}\mathbf{b}\\ \mathbf{f}_0 \\ \mathbf{f}_C\end{bmatrix}\right)\end{aligned} \qquad (30)$$

Then, the output equation is:

$$\mathbf{C}_S = -\mathbf{K}_{11}^{-1}\mathbf{K}_{12} \qquad (31)$$

and the feed through matrix is:

$$\mathbf{D}_S = -\mathbf{K}_{11}^{-1}[-\mathbf{K}_{b1} \quad \mathbf{I}_{11} \quad \mathbf{0}] \qquad (32)$$

# 5  Experimental validation

## 5.1  House description

Empirical validation of the detailed thermal simulation of buildings was undertaken in the framework of the IEA-EBC Annex 58 (Strachan et al. 2014). Measurements were performed on two typical detached houses operated by Fraunhofer Institute IBP, Holzkirchen, Germany (Figure 2), named N2 and O5. External walls of both houses are insulated, the windows are double glazed and equipped with external roller blinds. During the experiment, the roller blinds in all rooms were up, except the ones on the south façade. In case of N2 house, the South façade roller blinds were up for the entire duration of experiment, whereas in case of O5 house the South facade blinds were down during the initialisation period, i.e. during the first 7 days when the indoor temperature was constant, and up during all other periods. The ground floor of N2 house, consisting of 7 rooms, was modelled, with the cellar and the attic considered as known boundary condition. Doors between kitchen and living room, doorway and living room, bedroom and corridor were closed and sealed with tape. Doors connecting the corridor with the living room, bathroom and bedroom 2 were fully open. Infiltration rates were measured between the rooms and between indoor and outdoor. Mechanical ventilation provided an air flow rate of 120 m³/h in the living room, which was extracted equally in the bathroom and the bedroom 2.

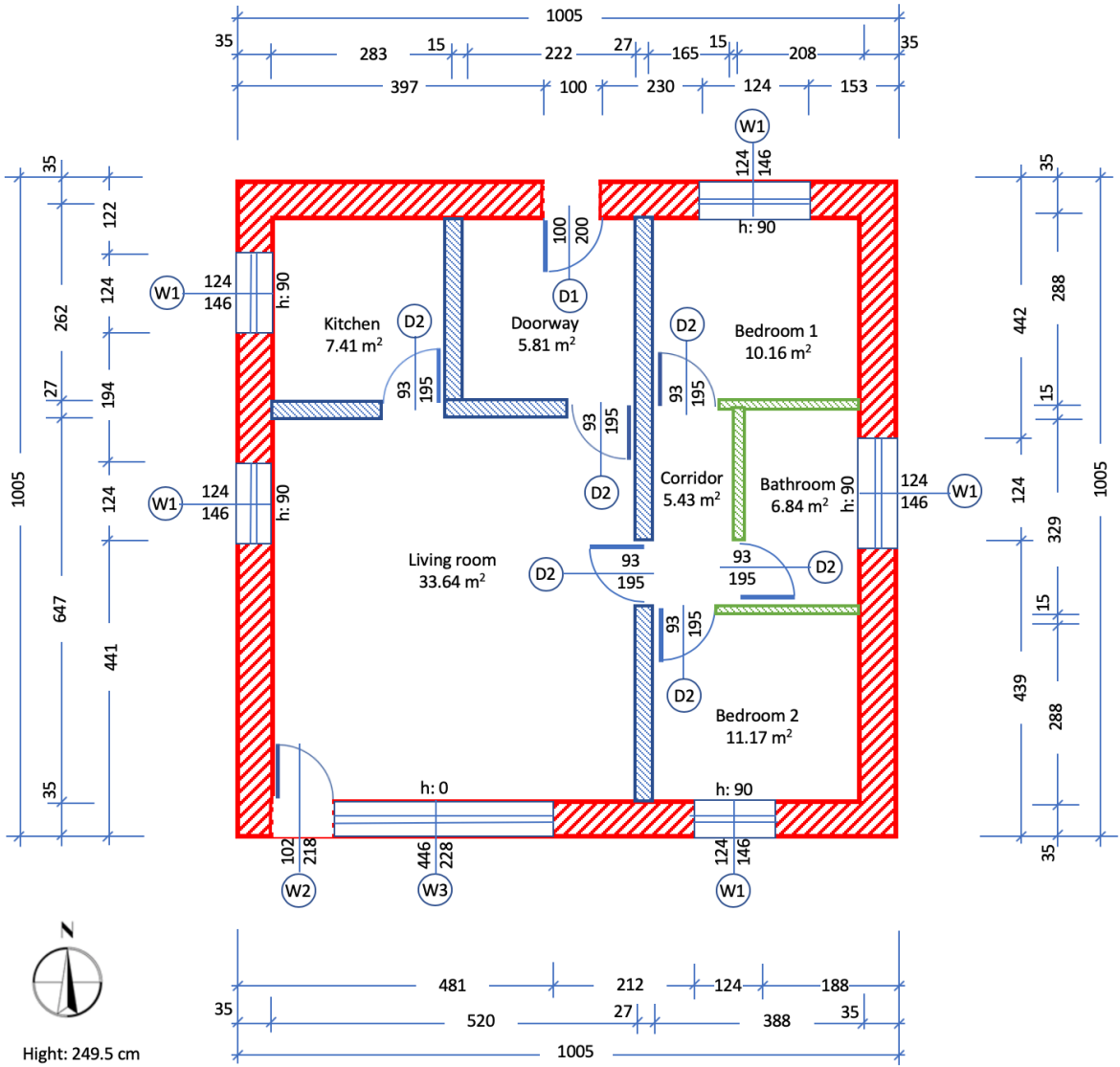

Figure 2 Plan of the house.

*Table 1 Dimensions of windows*

| Window type | Overall dimensions with roller blinds (m$^2$) | Glass dimensions (m$^2$) |
|---|---|---|
| W1 | 1.74 x 1.23 | 1.30 x 0.99 |
| W2 | 2.57 x 1.11 | 2.13 x 0.865 |
| W3 | 1.74 x 3.34 | 3 x 1.385 x 0.99 (3 panes) |

*Table 2 Walls and doors*

|   | Wall type | Layer | Thickness (m) | Conductivity (W/mK) | Dens. (kg/m$^3$) | Sp. heat (J/kg K) | Absorp. SW | Emiss. LW |
|---|---|---|---|---|---|---|---|---|
| 1 | External wall U = 0.2 | Ext. plaster | 0.01 | 0.80 | 1200 | 1000 | 0.23 | 0.90 |
|   |   | Insulation | 0.12 | 0.035 | 80 | 840 |   |   |
|   |   | Plaster | 0.01 | 1.00 | 1200 | 1000 |   |   |
|   |   | Brick | 0.20 | 0.22 | 800 | 1000 |   |   |
|   |   | Int. plaster | 0.01 | 1.00 | 1200 | 1000 | 0.17 | 0.90 |
| 2 | Internal wall | Plaster | 0.01 | 0.35 | 1200 | 1000 | 0.17 | 0.90 |
|   |   | Brick | 0.25 | 0.33 | 1000 | 1000 |   |   |
|   |   | Plaster | 0.01 | 0.35 | 1200 | 1000 | 0.17 | 0.90 |
| 3 | Internal wall | Plaster | 0.01 | 0.35 | 1200 | 1000 | 0.17 | 0.90 |
|   |   | Brick | 0.13 | 0.33 | 1000 | 1000 |   |   |
|   |   | Plaster | 0.01 | 0.35 | 1200 | 1000 | 0.17 | 0.90 |
| 4 | Ceiling U = 0.25 | Screed | 0.04 | 1.40 | 2000 | 2000 | 0.60 | 0.90 |
|   |   | Insulation | 0.04 | 0.04 | 80 | 840 |   |   |
|   |   | Concrete | 0.22 | 2.00 | 2400 | 1000 |   |   |
|   |   | Plaster | 0.01 | 1.00 | 1200 | 1000 |   |   |
|   |   | Insulation | 0.10 | 0.035 | 80 | 840 | 0.17 | 0.90 |
| 5 | Ground U = 0.32 | Concrete | 0.22 | 2.10 | 2400 | 1000 | 0.60 | 0.90 |
|   |   | Fill | 0.03 | 0.06 | 80 | 840 |   |   |
|   |   | Insulation | 0.03 | 0.025 | 80 | 840 |   |   |
|   |   | Panel | 0.03 | 0.023 | 80 | 840 |   |   |
|   |   | Screed | 0.06 | 1.40 | 2000 | 1000 | 0.60 | 0.90 |
| 6 | External door | Wood | 0.04 | 0.13 | 600 | 1000 | 0.60 | 0.90 |
| 6a | Internal door | Wood with glass | 0.04 | 0.13 | 600 | 1000 | 0.60 | 0.90 |
| 7 | Infiltration | 1.62 ACH |   |   |   |   |   |   |
| 8 | Ventilation | 60 m$^3$/hr |   |   |   |   |   |   |
| 9* | Pillar | Concrete |   | 0.5 | 2400 | 1000 | 0.60 | 0.90 |

*9 is numbered as 26

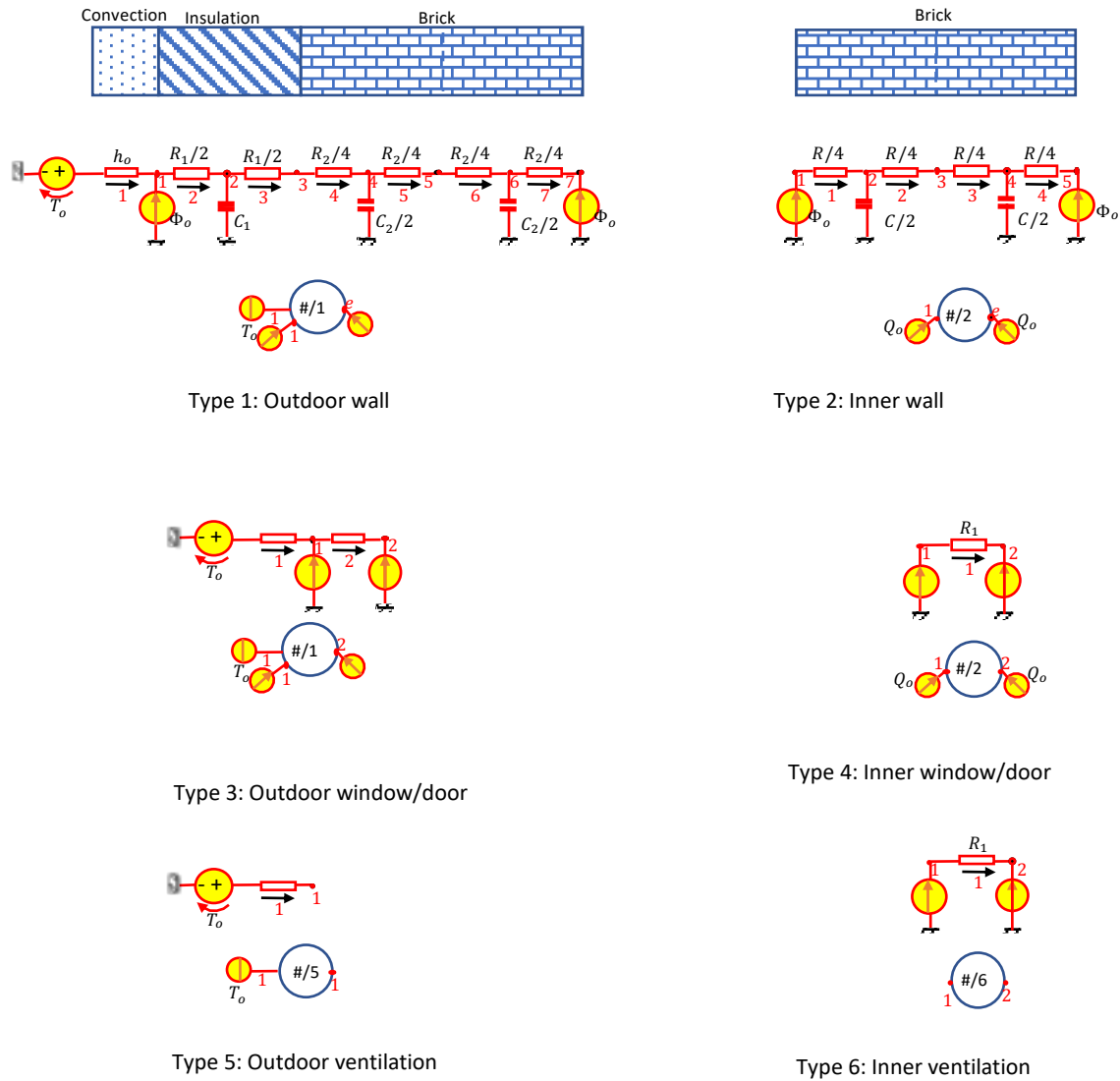

*Figure 3 Elementary models used in simulation*

### 5.2 Experimental protocol

The experiment was conducted for 41 days in summer 2013 starting with an initialisation period of 7 days followed by a period of 7 days with a heating set point of 30°C in all rooms. The constant temperature period was followed by a Randomly Ordered Logarithmic Binary Sequence of heat inputs (ROLBS) for 15 days. A 500 W heating power was supplied only in the living room during the ROLBS period. The time period for heat pulses ranged from 1 hour to 90 hours. The ROLBS period was followed by a re-initialisation period for 7 days with a heating set point of 25°C in all rooms. For the final 11 days of the experiment, no heat input was supplied and the temperature was allowed to float freely (Strachan et al. 2014).

### 5.3 Simulation results

The experimental house consists of seven thermal zones: kitchen, living room, doorway, bedroom 1, bathroom, corridor and bedroom 2. Internal temperature is simulated for living room (single zone), kitchen & living room (two zones), and entire twin house (seven zones), respectively (Figure 2).

A thermal circuit for each building element (wall, roof, window, door, etc.) is obtained by using techniques discussed in section 2 (Figure 3). The spatial discretisation of the building elements can be changed. The wall thermal capacity is located in the middle of each slice.

All thermal properties are considered constant in time (Tables 1 and 2). In simulation, the solar radiation on walls and windows is taken from measurements and from calculations; both approaches give very similar results.

Solar radiations entering the twin house through windows are calculated as a function of varying solar transmittance that changes with angle of incidence of solar radiation. According to the manufacturer, the heat from heaters is split in thirty percent radiation and seventy percent convection (Strachan et al. 2014). The same split of heat input from the heater is considered in simulation.

The thermal circuits generated for each zone are assembled as discussed in section 3. The state space model for the assembled circuits is generated as discussed in section 4.

The measured data has a time step of 10 minutes and the same time step of 10 minutes is used in simulation.

### 5.3.1 One-zone (living room)

The model of the living room consists of 13 thermal circuits (building elements). The numbering of each circuit and the position of each input to thermal circuit is shown in Figure 4. Thermal circuit number 13 represents the air node, which receives heat transferred from 12 thermal circuits that model the building elements.

The measured temperatures of all the zones external to the living room (including the walls of the kitchen, doorway, corridor and bedroom and the doors) form the boundary conditions.

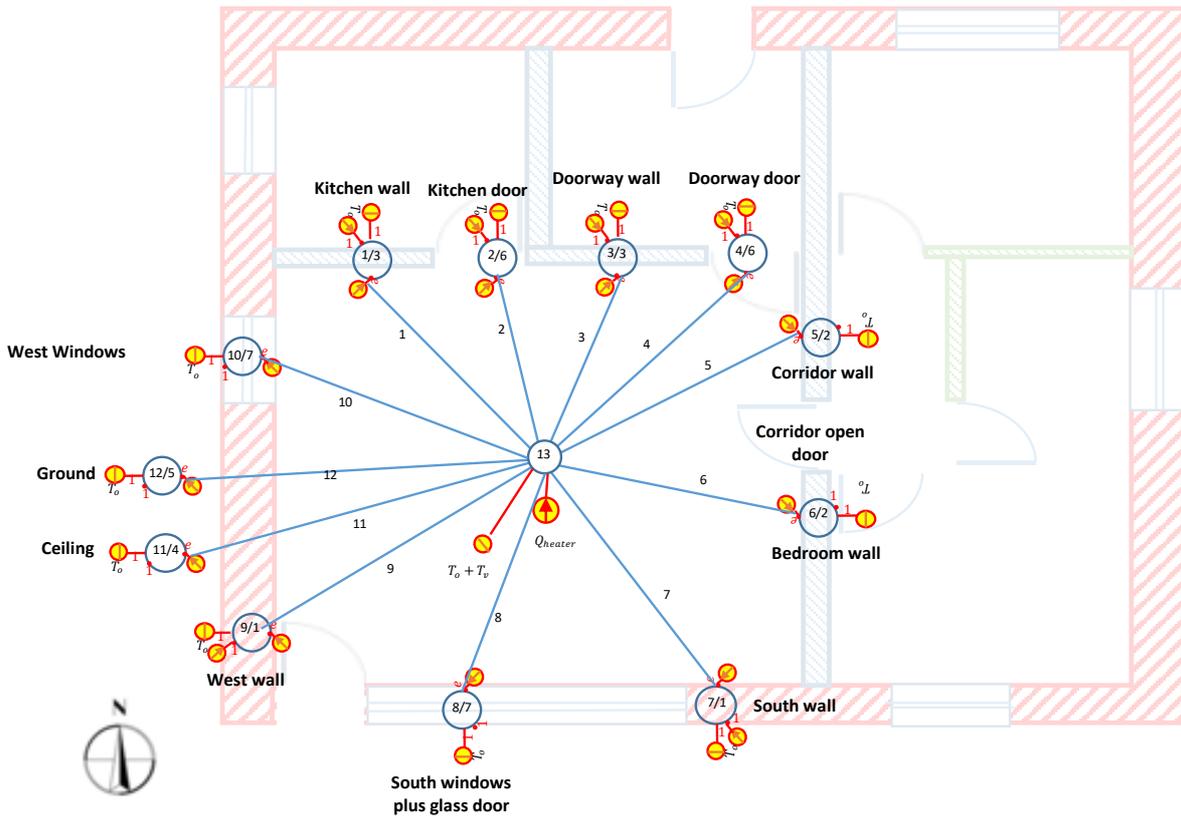

*Figure 4: Layout and thermal circuits for single zone (living room): 13 elementary models, 49 states*

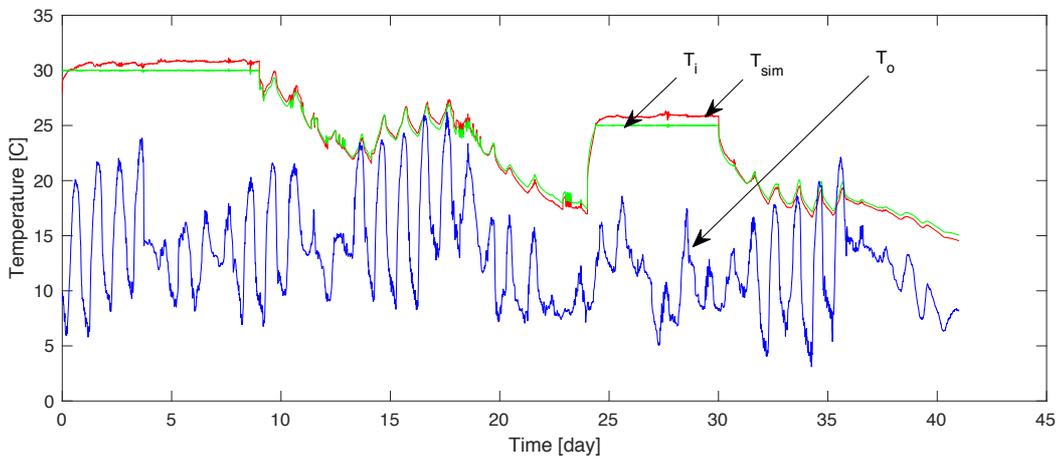

*Figure 5: Simulation results for living room; green line represents the measured internal temperature, $T_i$, the red line is the simulated temperature, $T_{sim}$, and the blue line presents the variation of the outdoor temperature, $T_o$, during the period of 41 days*

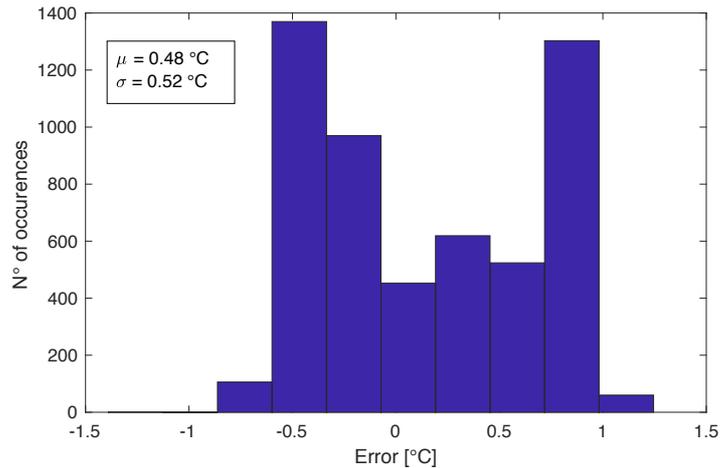

*Figure 6: Error histogram for living room*

The simulated internal temperature follows the measured temperature closely (Figure 5). The histogram of errors (Figure 6) shows that simulation errors lie well within $\pm 1°C$. The mean error is $\mu = 0.48\ °C$ and the standard deviation is $\sigma = 0.52\ °C$.

### 5.3.2 Two-zone model (living room and kitchen)

The kitchen and the living room were modelled together to validate the assembling and the state-space methodologies discussed in sections 3 and 4. The kitchen and the living room share an internal wall and a door. Although the shared door is sealed, there is infiltration of air between the two rooms, equivalent to 1/3$^{rd}$ of the infiltration between North and South zones. The ventilation supply duct passing through the kitchen is uninsulated and is responsible for loss of heat from the kitchen. The heat losses to ventilation duct are provided in the experimental data and are incorporated in the simulation.

The total number of thermal circuits for two zones is nineteen (Figure 7). All zones external to the kitchen and the living room are considered boundary conditions for the model. The number of state variables in the assembled model is 57.

Simulation of the indoor air temperature in the kitchen and the living room shows that the simulated air temperature lies within $+1.5$ and $-0.5\ °C$ with few outliers for kitchen and within $\pm 0.5°C$ for the living room, with a few outliers (Figure 8).

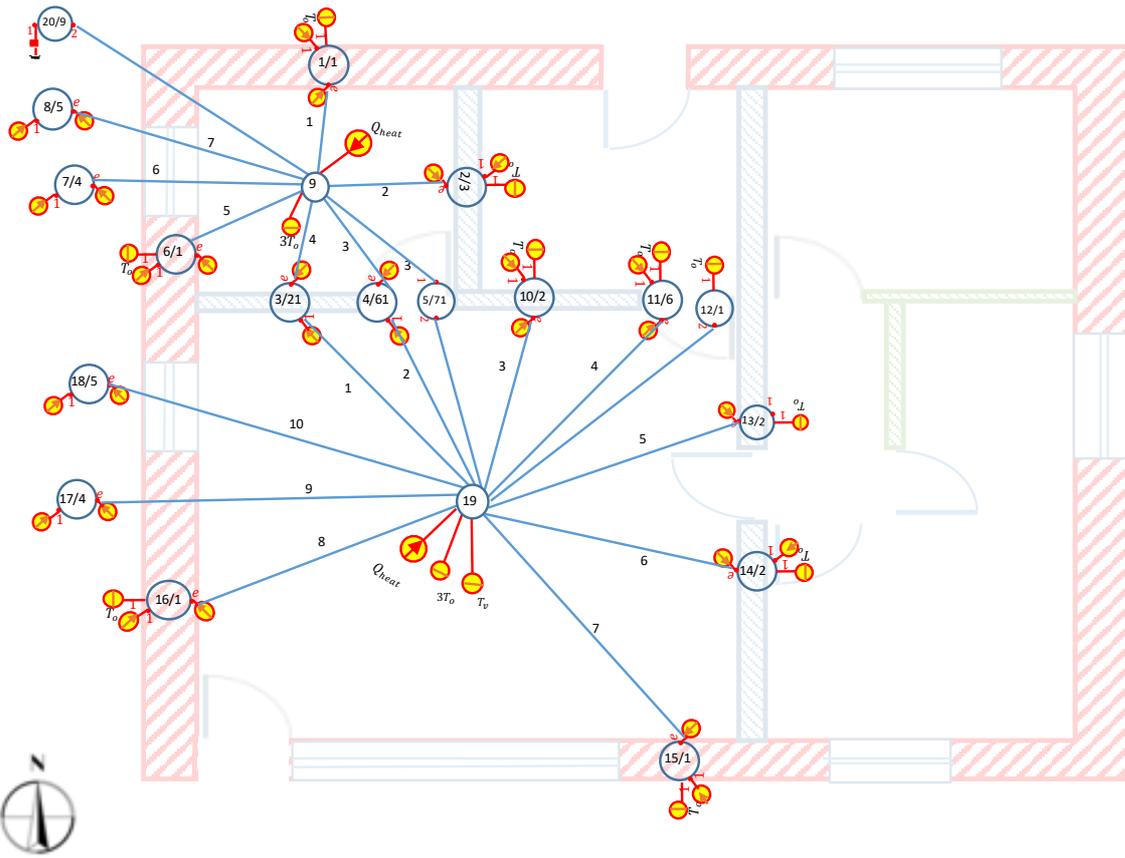

*Figure 7: Layout and thermal circuits for kitchen and living room: 2 thermal zones, 19 elementary model types, 57 states.*

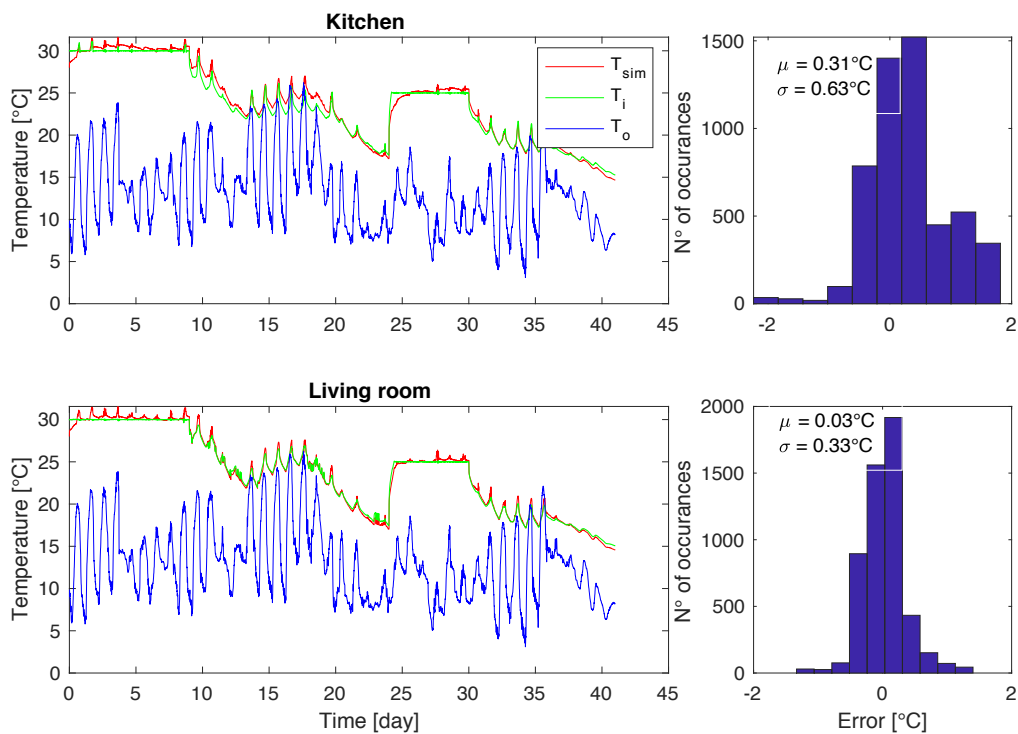

*Figure 8: Temperature simulation results and error histograms for kitchen and living room air temperature*

### 5.3.3 Seven thermal zones (whole house)

The model for the whole house consists of seven thermal zones that are modelled by assembling 56 thermal circuits (Figure 9). The number of states in the final state-space model is 109. The simulation error increases with the number of zones, which is explained by the errors induced in the values of boundary conditions (Figure 10). The simulation results show that for every zone three quantiles of simulated temperature lie within $\pm 1°C$.

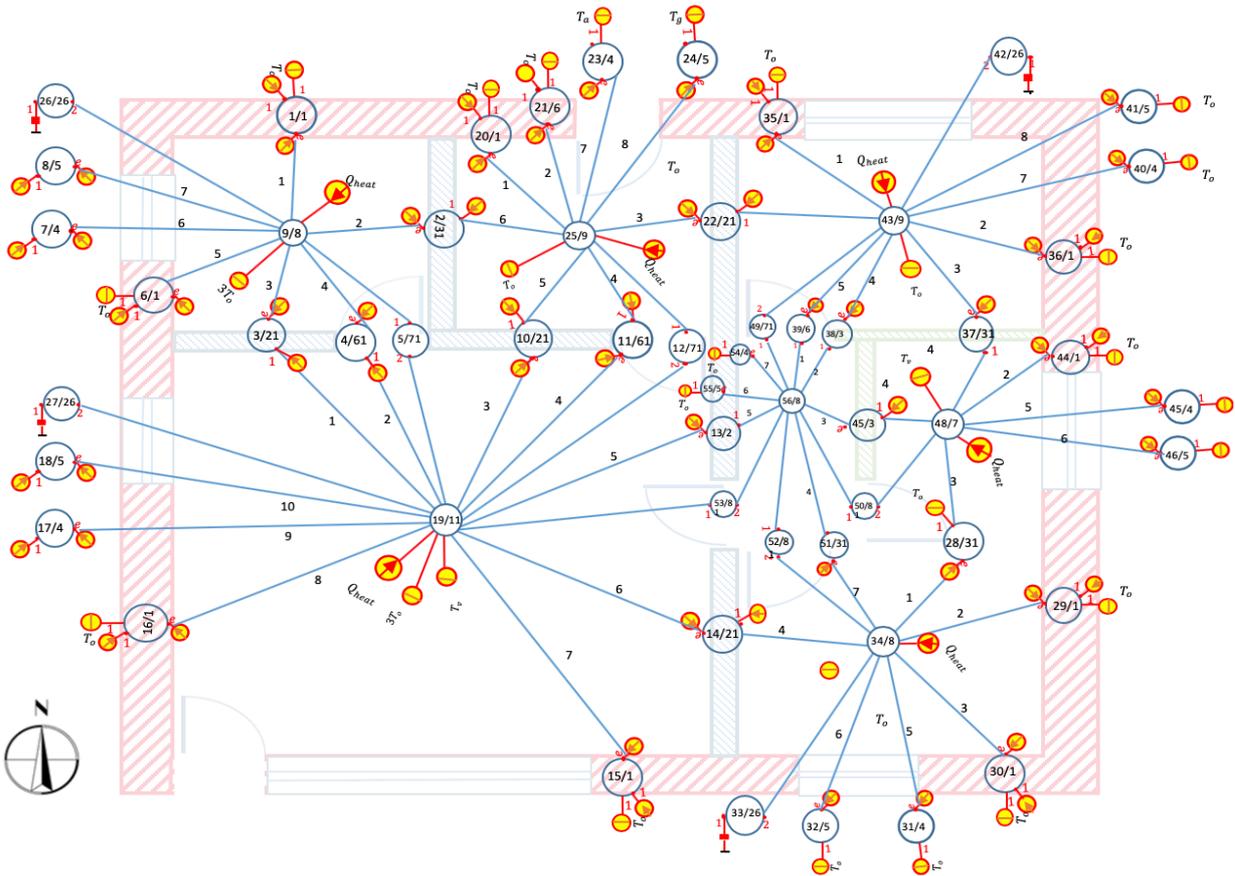

*Figure 9 Assembled model of a house: 7 thermal zones, 56 elementary model types, 109 states.*

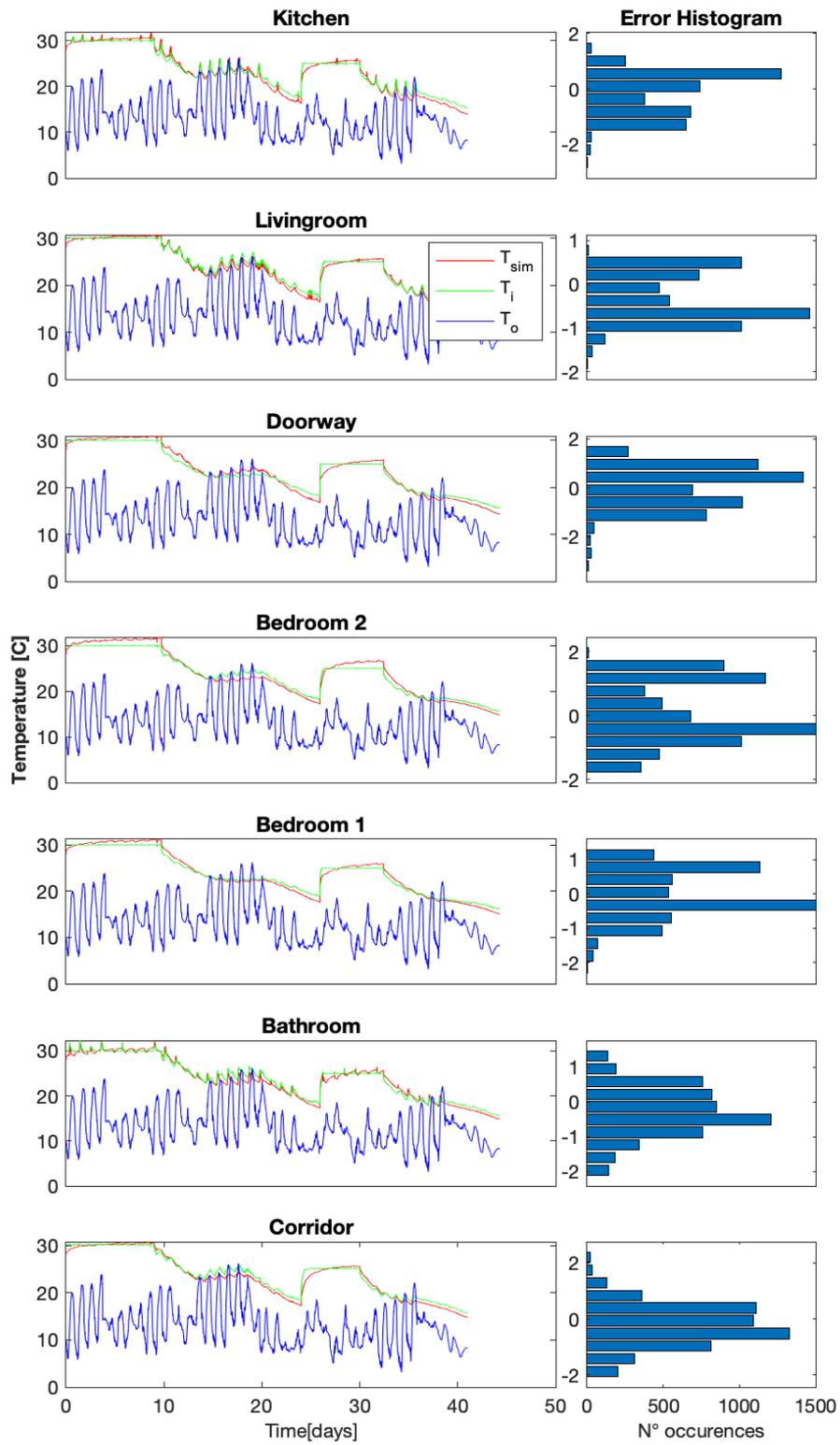

*Figure 10: Indoor air temperatures and error histograms for whole twin house (seven zones)*

# 6 Conclusions

State-space representation can be effectively obtained from thermal circuits, even for very large models. This is specifically suited to detailed thermal models of buildings. The state-space representation, although linear, can also be used for non-linear models if the linearity is considered for a time step. State-space models are completely equivalent to thermal circuits from which they were obtained. The assembling of elementary thermal circuits allows us to obtain only one thermal circuit for the whole building; therefore, the state-space model can model in detail a whole building. Since assembling can be used to create large models from individual elements, it can find applications in the emerging technologies of Building Information Modelling (BIM). Obtaining state-space representation from very large thermal networks can have applications in model order reduction and the synthesis of control algorithms for complex buildings.